\documentclass[aps,prb,reprint,superscriptaddress]{revtex4-1}

\usepackage[usenames]{color}
\usepackage{colortbl}
\usepackage{amsmath}
\usepackage{graphicx}
\usepackage[right]{lineno}
\usepackage{dcolumn}

\begin{document}





\title{Optical spectroscopy of single beryllium acceptors in GaAs/AlGaAs quantum well}

\author{P.V. Petrov}
\email{pavel.petrov@gmail.com}
\affiliation{Ioffe Institute, St. Petersburg, Russian Federation}

\author{I.A. Kokurin}
\affiliation{Ioffe Institute, St. Petersburg, Russian Federation}
\affiliation{Mordovia State University, Saransk, Russian Federation}

\author{G.V. Klimko}
\author{S.V. Ivanov}
\author{Yu.L. Iv\'anov}
\affiliation{Ioffe Institute, St. Petersburg, Russian Federation}

\author{P.M. Koenraad}
\author{A.Yu. Silov}
\affiliation{Department of Applied Physics, Eindhoven University of Technology, The Netherlands}

\author{N.S. Averkiev}
\affiliation{Ioffe Institute, St. Petersburg, Russian Federation}




\begin{abstract}
We carry out micro-photoluminescence measurements of an acceptor-bound exciton (${\rm A^0X}$) recombination
in the applied magnetic field with a single impurity resolution. In order to describe the obtained spectra we develop a theoretical
model taking into account a quantum well (QW) confinement, an electron-hole and hole-hole exchange interaction.
By means of fitting the measured data with the model we are able to study the fine structure of individual acceptors
inside the QW. The good agreement between our experiments and the model indicates that
we observe single acceptors in a pure 2D environment which states are unstrained in the QW plain.
\end{abstract}


\maketitle

\section{Introduction}
Studies of single impurities in solids is one of the most rapidly developing fields
of experimental physics in recent years~\cite{koenraad2011single, PhysRevLett.93.207403, kudelski2007optically}.
Such experiments are attractive since it makes possible to verify the fundamental theoretical approaches
that were based on macroscopic measurements. In the field of applied science
a device based on single impurities is the ultimate limit of electronics miniaturization.
At present, two main techniques are exploited in order to reach a single impurity resolution:
scanning tunneling microscopy~\cite{PhysRevLett.92.216806} and micro-photoluminescence. On the one hand the optical methods
have advantage over tunneling measurements due to absence of the surface influence.
On the other hand the resolution of optical measurements is fundamentally restricted by the diffraction limit.
The photon wavelength has to be smaller than the average distance between impurities.
The band gap of typical semiconductors is about 1$\,$eV, and therefore the corresponding
doping concentration should not exceed $10^{12}\,$cm$^{-3}$ for 3D or $10^{8}\,$cm$^{-2}$ for 2D systems. 
At present, the spectroscopy of single semiconductor nanostructures such as quantum dots (QD) is well developed~\cite{PhysRevLett.74.4043}.
The obvious approach is to dope a single QD with an impurity atom.
Experiments of this kind were realized for CdTe~\cite{PhysRevLett.93.207403} and InAs~\cite{kudelski2007optically} QDs doped with Mn.
However, an interpretation of experimental results in QD systems is hampered by the fact that
such parameters as a dot size, shape, chemical composition as well as an impurity position 
inside the QD are randomly distributed across QD ensemble. 
It makes necessary to use a lot of additional parameters in the theoretical description
of experimental results~\cite{PhysRevB.76.045331, krebs2009magnetic}.

In the present article, we study a narrow GaAs/AlGaAs QW doped with beryllium in order to optically explore single impurities.
Usually, the single emitters in such systems are studied via sub-micron apertures or mesa structures formed on a sample surface~\cite{gammon1996fine, PhysRevLett.89.177403}.
To reach the single impurity resolution here, we do not use any preprocessing of the samples but optimize the doping process instead.
The smallest controllable sheet impurity density in our experiments is about $10^{10}\,$cm$^{-2}$. This number does not meet
the diffraction limit condition, but nonetheless can serve a purpose in the same way as was first realized in 
the spectroscopy of single organic molecules~\cite{PhysRevLett.62.2535}.
The point is to put emitters in a media that randomly changes the emitters energy
and to employ a spectral resolution in addition to spatial one.
It is well known that fluctuations in a QW width lead
to an inhomogeneous spectral broadening of the exciton photoluminescence due to significant variations of the effective bandgap~\cite{Weisbuch1981709}.
Assuming that the energy broadening corresponds to a Gaussian shape of photoluminescence line,
let us consider the low-energy tail of spectrum.
For the Gaussian distribution a probability that the transition energy
is in the range between 2 and 3 standard deviations from the distribution maximum is about 1\%.
Therefore we can reach the necessary small sheet density $10^{8}\,$cm$^{-2}$ of impurity related single optical emitters,
if we examine a lower-energy tail in the photoluminescence of an inhomogeneously broadened ensemble of the impurities.

The interface roughness leads to lateral asymmetry and affects the energy structure of excitons~\cite{gammon1996fine, Goupalov19981205}.
But if a radius of an impurity-bound exciton is smaller than a scale of the roughness,
we can neclect the lateral asymmetry and consider such an exciton in a pure 2D environment.
This allows us to significantly reduce the number of fitting parameters in comparison with
the case of doped QDs~\cite{PhysRevB.76.045331, krebs2009magnetic}.

This article is organized as follows: we describe the sample growth, the characterization procedure and micro-photoluminescence
data in Sec. II. In Sec. III we present a theoretical model of the acceptor-bound exciton which includes the QW confinement.
A comparison of theoretical calculation with obtained and previously published experimental data is discussed in Sec. IV.

\section{Experiment}
We grew by molecular beam epitaxy three GaAs/Al$_{\rm x}$Ga$_{\operatorname{1-x}}$As
QW structures with the QW width of 3.7$\,$nm and the Al content in the
barriers x=0.25. The samples were doped with Be acceptors inside the QW.
They have similar design and differ only in the Be doping mode and sheet
impurity density, as shown in Table 1. In two samples the QW is
$\delta$-doped in the middle, while in the third we use uniform doping of the
QW with 1-ML-thick undoped spacers at both interfaces.
We adjusted the barrier height in order to ensure an effective band-to-band absorption of a pumping light inside the barriers.

\begin{table*}
\begin{ruledtabular}
\begin{tabular}{p{1.0cm}p{2.8cm}p{1.0cm}p{1.65cm}cp{5.7cm}cp{1.65cm}c}
\mbox{Sample} \mbox{No.} &\hspace{0.7cm}Substrate  \mbox{$\rm T_{sub}=560-580\rm ^o$C} & GaAs buffer &\mbox{Al$_{\rm x}$Ga$_{\operatorname{1-x}}$As} \mbox{\hspace{0.3cm}x=0.25} & GaAs &\hspace{2.3cm}GaAs:Be & GaAs &\mbox{Al$_{\rm x}$Ga$_{\operatorname{1-x}}$As} \mbox{\hspace{0.3cm}x=0.25} & GaAs \\
\colrule
\hspace{0.15cm}S1 &undoped GaAs & 0.25$\,\mu$m & \hspace{0.3cm}100$\,$nm & 7$\,$ML &\rule{0pt}{10pt}\hspace{0.34cm}$\delta\,$3'', $\rm T_{Be}=660^o$C, $N_s=5\cdot 10^9\,$cm$^{-2}$ & 6$\,$ML & \hspace{0.3cm}100$\,$nm & 20$\,$nm \\
\hspace{0.15cm}S2 &p-type GaAs & 0.25$\,\mu$m & \hspace{0.3cm}100$\,$nm & 7$\,$ML &\rule{0pt}{10pt}\hspace{0.34cm}$\delta\,$3'', $\rm T_{Be}=690^o$C, $N_s=5\cdot 10^{10}\,$cm$^{-2}$ & 6$\,$ML & \hspace{0.3cm}100$\,$nm & 20$\,$nm \\
\hspace{0.15cm}S3 &undoped GaAs & 0.25$\,\mu$m & \hspace{0.3cm}100$\,$nm & 1$\,$ML &\rule{0pt}{10pt}11$\,$ML, $\rm T_{Be}=660^o$C, $N_s=4\cdot 10^{10}\,$cm$^{-2}$ & 1$\,$ML & \hspace{0.3cm}100$\,$nm & 20$\,$nm \\
\end{tabular}
\end{ruledtabular}
\caption{Sample parameters: $\rm T_{sub}$ and $\rm T_{Be}$ is the temperature of the substrate and Be source correspondingly,
$\delta\,$3'' stands for 3 seconds of $\delta$-doping, 1$\,$ML means one half of the lattice constant,
$N_s$ is an expected acceptor sheet concentration inferred from the beryllium source calibration.
 }
\end{table*}

We use macro-photoluminescence measurements at $4.2\,$K in order to characterize the grown samples
and to establish the presence of beryllium inside the QW.
The samples were pumped with a 660$\,$nm diode laser via an optical fiber with the cross section of 0.1$\,$mm$^2$,
macro-luminescence spectra were collected through the same fiber.
We expect that the beryllium dopants reveal itself as an additional low-energy broadening of
the QW related photoluminescence line due to an appearance of the acceptor-bound excitons \cite{PhysRevB.40.10021}
Samples S2 and S3 with more intensive doping indeed demonstrate the expected low-energy broadening
as it shown in Fig.~1. The micro-photoluminescence measurements show that these tails consist of numerous narrow
lines. In order to distinguish the micro-photoluminescence lines due to the acceptor-bound excitons
from the lines of different origin, we use the rich energy structure of ${\rm A^0X}$ complex~\cite{PhysRevB.47.15675}
as a spectral fingerprint.
\begin{figure}[b]
\includegraphics[width=8cm]{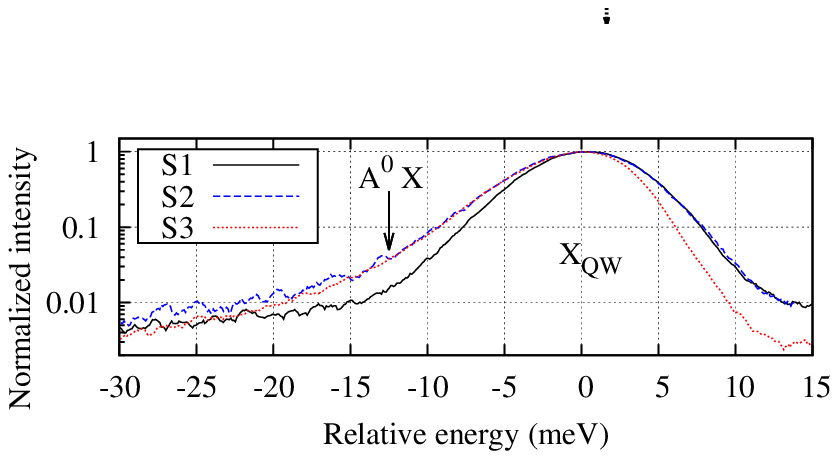}
\caption{\label{fig01} (color online) Normalized macro-photoluminescence spectra of the studied samples. 
Spectra are centered at photoluminescence maxima for comparison; the pump density is 10$\,$W/cm$^2$.}
\end{figure}
\begin{figure}[b]
\includegraphics[width=8cm]{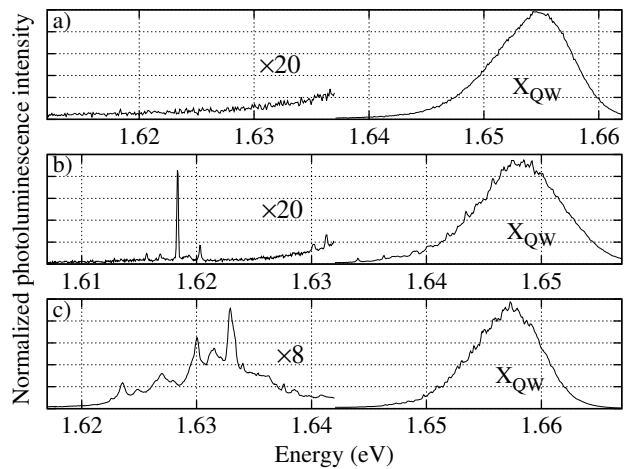}
\caption{\label{fig02} Three typical micro-photoluminescence spectra of the studied samples that were measured
at different spots on the samples. The spectrum in panel {\bf a)} contains only broad line of excitons (${\rm X_{QW}}$) which recombine inside
the QW; no extra features present at the low-energy tail.
Narrow well-resolved lines are present on the low-energy side of the spectrum of type {\bf b)}.
The panel {\bf c)} depicts spectrum which contains numerous overlapping lines at the region of interest.}
\end{figure}

We carry out micro-photoluminescence measurements at 5$\,$K on the setup with about 1$\,\mu$m spatial
and $\sim 60\,\mu$eV energy resolution using HeNe laser as a pump source.
Figure~2 shows three characteristic spectra which were measured at different spots on the surface of our samples.
The strong photoluminescence line at 1.65--1.66$\,$eV corresponds to recombination of excitons 
inside the QW (${\rm X_{QW}}$) and it looks the same for all the samples, while the low-energy side
of the spectrum presents a wide variety of results.
We observe mostly spectra of type {\bf a)} on the sample S1 with the lowest doping concentration.
There are no evidences of impurities at the low-energy tail in panel {\bf a)}.
The sample S3 shows a strong non-uniformity across the surface: most of the sample surface corresponds to
the type {\bf a)} while less than 10\% of the surface gives the spectra of type {\bf c)}.
The spectrum in panel {\bf c)} contains an impurity related luminescence at the low-energy side,
but the luminescence lines are quite broad and overlap each other.
The sample S2 is the one most suitable for micro-photoluminescence measurements.
We observe the spectra of type {\bf b)} with strong narrow lines at the low-energy tail
of the sample S2 photoluminescence. Below in the text we discuss results
which were obtained on this sample.

In trying to find out if there are any distinctive peculiarities in the micro-photoluminescence spectra,
we carry out measurements in the applied magnetic field in Faraday geometry.
Amongst manifold combinations of luminescence lines we observe a kind of repeating pattern in polarized photoluminescence spectra.
It consists of one strong single luminescence line and two weak adjacent satellites which are split
by magnetic field in doublets denoted as (1), (2) and (3) in Fig.~3.
It is noteworthy that a Zeeman splitting of the satellites is 1.5--2 times stronger than a splitting of the main line.
In order to give a reliable interpretation of the results we develop a theoretical model of an acceptor-bound exciton
in which we take into account an interparticle exchange~\cite{PhysRevB.11.5002, pikus1980fine}, the QW confinement~\cite{PhysRevB.47.15675}
and the magnetic field.

\begin{figure}[!tbp]
\includegraphics[width=8cm]{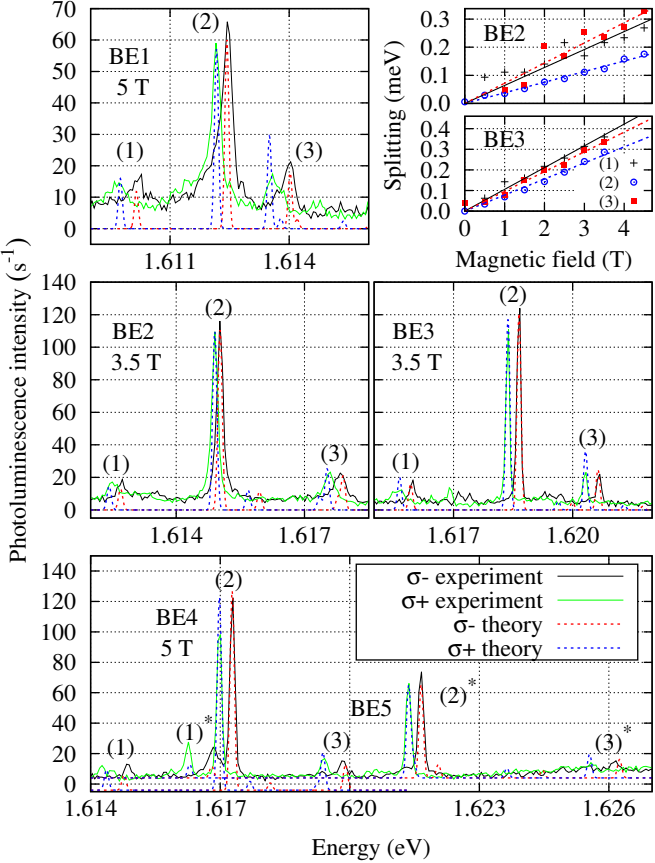}
\caption{\label{fig03} (color online) Experimental (solid line) and theoretical (dashed line) polarized micro-photoluminescence
spectra of single acceptor-bound excitons in the applied magnetic field.
Small panels display the Zeeman splitting. Symbols are experimental data
for doublets (1), (2) and (3), solid lines are the linear least-squares fitting.}
\end{figure}

\section{Theory}
In order to obtain an energy structure of acceptor-bound exciton ${\rm A^0X}$ inside a QW,
we use a model Hamiltonian:
\begin{center}
\begin{equation}
\begin{split}
H=-\Delta^{{\rm hh}}\mathbf{J}_1\cdot\mathbf{J}_2-\Delta^{{\rm eh}}{\bf
S}\cdot({\bf J}_1+{\bf J}_2)+ \\
+\frac{\Delta^{{\rm qw}}}{2}\left(J_{1z}^2+J_{2z}^2-\frac52\right),
\end{split}
\end{equation}
\end{center}
where $\Delta^{{\rm hh}}$ and $\Delta^{{\rm eh}}$ are the hole-hole and the electron-hole exchange energies, respectively,
and $\Delta^{{\rm qw}}$ is a splitting of the localized hole state due to a QW confinement.
Here ${\bf J}_i$ ($i=1,2$) and  $\bf S$ stand for an angular momentum of the holes and the electron, respectively.
We use a spherical model of the localized hole states \cite{PhysRevB.8.2697} and consider only the ground state with momentum $J=3/2$.
The wavefunction of two indistinguishable holes must be antisymmetric,
therefore only states with total angular momentum $J=0,2$ are present.
In diamond-like semiconductors the hole-hole exchange is a ferromagnetic interaction
($\Delta^{{\rm hh}}>0$), therefore a state with the largest total angular momentum is the ground one~\cite{averkiev1993many}.
The electron-hole exchange interaction between these two holes and the electron with $S=1/2$ leads to the emergence of a three-particle complex
with total angular momentum $F=1/2, 3/2, 5/2$.
This interaction is also ferromagnetic ($\Delta^{{\rm eh}}>0$) in GaAs/AlGaAs QWs~\cite{PhysRevLett.86.5176};
it means that the ``dark'' state of the free exciton is the ground state.
An energy splitting due to the QW confinement is negative ($\Delta^{{\rm qw}}<0$)
which corresponds to $J_z=\pm 3/2$ as a ground hole state.
The order of levels in the bulk ${\rm A^0X}$ complex ($\Delta^{{\rm qw}}=0$) depends on the ratio between $\Delta^{{\rm hh}}$ and $\Delta^{{\rm eh}}$.
We obtain all the three-particle wave functions $\Psi^{5/2}_{\pm 5/2}$, $\Psi^{5/2}_{\pm 3/2}$, $\Psi^{5/2}_{\pm 1/2}$,
$\Psi^{3/2}_{\pm 3/2}$, $\Psi^{3/2}_{\pm 1/2}$ and $\Psi^{1/2}_{\pm 1/2}$ analitically using the usual procedure of angular momentum coupling \cite{landau1965quantum}.
Here the upper index is a full angular momentum of the state while the lower one is its projection.
The energy levels of ${\rm A^0X}$ complex are given by solution of Schr\"{o}dinger equation with Hamiltonian~(1)
at $\Delta^{{\rm qw}}=0$.

\begin{gather}
E_{1/2}=\frac{15}{4}\Delta^{{\rm hh}},~~~
E_{3/2}=\frac34\Delta^{{\rm hh}}+\frac32\Delta^{{\rm eh}}, \nonumber \\
E_{5/2}=\frac34\Delta^{{\rm hh}}-\Delta^{{\rm eh}}
\end{gather}

\begin{figure}[t]
\includegraphics[width=8cm]{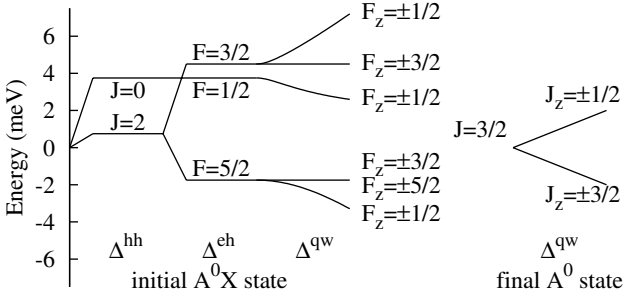}
\caption{\label{fig04} The scheme of energy levels for initial ${\rm A^0X}$ and final ${\rm A^0}$ states
in presence of exchange interaction and QW-confinement.}
\end{figure}

The QW potential leads to the mixing of levels with angular momentum projection $F_z=\pm 1/2$ keeping other levels constant.
Using $\Psi^{1/2}_{\pm 1/2}$, $\Psi^{3/2}_{\pm 1/2}$, $\Psi^{5/2}_{\pm 1/2}$ functions as a basis we can write the Hamiltonian
for three states with $F_z=\pm 1/2$:
\arraycolsep=0.3em
\begin{equation}
H=\left(\begin{array}{ccc}
\frac{15}{4}\Delta^{{\rm hh}}&-\sqrt{\frac 25}\Delta^{{\rm qw}}&\sqrt{\frac 35}\Delta^{{\rm qw}}\\
-\sqrt{\frac25}\Delta^{{\rm qw}}&\frac34\Delta^{{\rm hh}}+\frac32\Delta^{{\rm eh}}&0\\
\sqrt{\frac 35}\Delta^{{\rm qw}}&0&\frac34\Delta^{{\rm hh}}-\Delta^{{\rm eh}}
\end{array}\right)
\end{equation}
We obtain the energy levels $E_i$~($i=1,2,3$) and the corresponding
wave functions $\Psi^{(i)}_{\pm 1/2}=a^i_{1/2}\Psi^{1/2}_{\pm 1/2}+a^i_{3/2}\Psi^{3/2}_{\pm 1/2}+a^i_{5/2}\Psi^{5/2}_{\pm 1/2}$
as a solution of the Hamiltonian~(3) eigenvalue problem.
Figure~\ref{fig04} depicts the obtained energy scheme of ${\rm A^0X}$ complex.

Assuming that Zeeman energy is much smaller than all the energy parameters of the system,
we find the Zeeman splitting of ${\rm A^0X}$ levels in the
first order of perturbation theory. For simplicity sake
we take the electron g-factor $g_{e}=0$, which is true
for narrow GaAs/AlGaAs QWs~\cite{PhysRevB.75.245302}.
Zeeman splitting is described by the Hamiltonian:
\begin{equation}
H_\mathrm{Z}=\mu_Bg_{hi}B(J_{1z}+J_{2z}),
\end{equation}
where $g_{hi}$ is a hole g-factor of the initial ${\rm A^0X}$ state.
Let us obtain all the g-factors of the acceptor-bound exciton states
normalized to the angular momentum 1/2:
\[
g^{3/2}_{3/2}=\frac{18}{5}g_{hi},\qquad g^{5/2}_{5/2}=4g_{hi},\qquad
g^{5/2}_{3/2}=\frac{12}{5}g_{hi}.
\]
The g-factors of mixed states depend on the eigenvector coefficients $a^i_j$ ($i=1,2,3$,
$j=1/2,3/2,5/2$):
\begin{equation}
\frac{g_i}{g_{hi}}=\frac25\left(\sqrt{3}a^i_{3/2}+\sqrt{2}a^i_{5/2}\right)^2.
\end{equation}
The final state after the ${\rm A^0X}$ recombination is a neutral acceptor ${\rm A^0}$.
The final state is also split by the QW potential~\cite{PhysRevB.63.195317} with the same $\Delta^{{\rm qw}}$:
\begin{equation}
H=\frac{\Delta^{{\rm qw}}}{2}\left(J_z^2-\frac54\right).
\end{equation}
The four nondegenerate states of ${\rm A^0}$ produced by a magnetic field are:
\begin{gather}
E_{\pm 3/2}=\frac{\Delta^{{\rm qw}}}{2}\pm\frac32g_{hf}\mu_BB,\\
E_{\pm 1/2}=-\frac{\Delta^{{\rm qw}}}{2}\pm\frac12g_{hf}\mu_BB,
\end{gather}
where $g_{hf}$ is a g-factor of the final ${\rm A^0}$ state.

Knowing the energy of the initial $E_i$ and final $E_f$ states we can establish all transition energies as
\begin{equation}
\hbar\omega = E_g + E_i - E_f,
\end{equation}
where $E_g$ is an effective band gap including all the confinement shifts and the exciton binding energy.
In order to obtain oscillator strengths and polarizations of the transitions we use the usual selection rules
combined with Clebsch-Gordan coefficients that couple spins of ${\rm A^0X}$ complex.

\section{Discussion} 
Figure~\ref{fig03} shows a set of the circularly polarized micro-photoluminescence spectra which were measured
at different spots on the sample in applied magnetic field.
As mentioned above, all the spectra match a repetitive pattern: a strong line with two accompanying satellites.
According to our model the strongest photoluminescence lines which are denoted as (1), (2) and (3) correspond to
transitions $\Psi^{5/2}_{\pm 3/2} \rightarrow J_z=\pm 1/2$, $\Psi^{(1)}_{\pm 1/2} \rightarrow J_z=\pm3/2$
and $\Psi^{5/2}_{\pm 5/2} \rightarrow J_z=\pm3/2$, respectively. The most intense line (2) originated from the ground $\Psi^{(1)}_{\pm 1/2}$ state
of ${\rm A^0X}$ complex while satellites are due to the subsequent degenerate $\Psi^{5/2}_{\pm 3/2}, \Psi^{5/2}_{\pm 5/2}$ state.
The energy spacing between (1) and (3) lines is equal to $\Delta^{{\rm qw}}$ parameter of our model.
Sets of the fitting parameters of all spectra presented in Fig.~\ref{fig03} are compiled in the table~II under the labels BE1--BE5.

\begin{table}[!tbp]
\begin{center}
\begin{ruledtabular}
\begin{tabular}{cdddddc}
\mbox{No.} & \multicolumn{1}{c}{$\Delta^{{\rm hh}}$(meV)} &
\multicolumn{1}{c}{ $\Delta^{{\rm eh}}$(meV) }&
\multicolumn{1}{c}{ $\Delta^{{\rm qw}}$(meV) }&
\multicolumn{1}{c}{ $g_{hi}$ }&
\multicolumn{1}{c}{ $g_{hf}$ }&
 $T$(K)\\
\colrule
\rule{0pt}{10pt}BE1 & 1.0 & 2.3 & -3.81 & 0.75 & 0.4 & 20 \\ 
BE2 & 0.8 & 2.0 & -5.15 & 0.65 & 0.25 & 30 \\ 
BE3 & 1.2 & 2.3 & -4.70 & 0.8 & 0.55 & 20 \\ 
BE4 & 1.0 & 2.0 & -5.07 & 0.75 & 0.4 & 20 \\ 
BE5 & 2.5 & 3.0 & -9.35 & 0.9 & 0.4 & 50 \\ 
BE6 & 0.37 & 2.27 & -4.47 & & & 60 \\ 
\end{tabular}
\end{ruledtabular}
\end{center}
\caption{Fitting parameters used for calculation of spectra in Fig.~\ref{fig03} and Fig.~\ref{fig05}.}
\end{table}

\begin{figure}[!t]
\includegraphics[width=8cm]{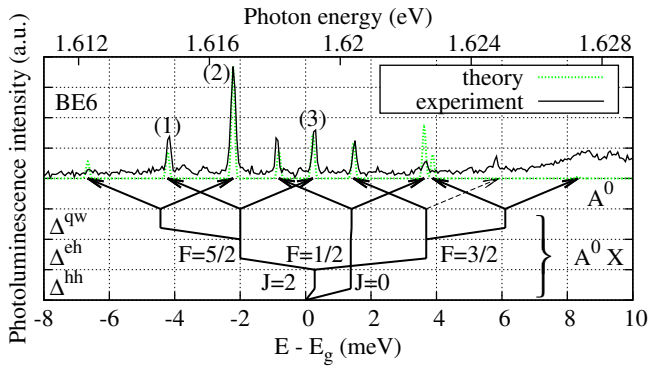}
\caption{\label{fig05} (color online) Experimental (solid line) and theoretical (dashed line) micro-photoluminescence spectra of an exciton
bound to the single beryllium acceptor. 
The level scheme depicts splitting of the initial ${\rm A^0X}$ state due to the hole-hole $\Delta^{{\rm hh}}$,
the electron-hole $\Delta^{{\rm eh}}$ exchange and the QW confinement potential $\Delta^{{\rm qw}}$;
arrows indicate a splitting in the final ${\rm A^0}$ state.
The dashed arrow marks a forbidden transition, which becomes available if we take into account
the cubic anisotropy of the crystal.}
\end{figure}

Another characteristic feature of our model comes from a fact that a radiative recombination
of acceptor-bound exciton occurs via transitions between a multiplet of initials states and only two available final states. 
It means that a few pairs of the spectral line with an equal spacing of $\Delta^{{\rm qw}}$ can be present
in the spectrum of the acceptor-bound exciton recombination.
Figure~5 depicts an experimental spectrum which contains three pairs with the similar energy splitting of $\Delta^{{\rm qw}}=4.47\,$meV.
Assuming that the most intensive line in the spectrum is a $\Psi^{(1)}_{\pm 1/2} \rightarrow J_z=\pm3/2$ transition
we successfully describe all other optical transitions
using our 3-parameter fit at zero magnetic field. In order to fit the transition intensities
we also take into account an equilibrium probability $\sim\exp(-E_i/kT)$  to find the ${\rm A^0X}$~complex in a certain initial state
using an effective bath temperature $T$ as a fourth parameter. The corresponding fit parameters are listed in the table~II denoted by BE6.

Let us compare the values of parameters with published results of other experiments.
Using equations~(2) we extract the values $\Delta^{{\rm hh}}_{3D}=0.11\,$meV and $\Delta^{{\rm eh}}_{3D}=0.06\,$meV from ${\rm A^0X}$ photoluminescence data
obtained on the bulk GaAs material~\cite{PhysRevB.11.5002}. It is well known that a quantum confinement
significantly enhances the electron-hole exchange in nanostructures~\cite{Goupalov1998393, PhysRevB.47.4569},
therefore our fitted values of exchange parameters $\Delta^{{\rm hh}}$ and $\Delta^{{\rm eh}}$ seem reasonable.
The typical effective temperature of recombining excitons
is about $20\,$K in narrow GaAs/AlGaAs QWs~\cite{PhysRevB.39.3419} in accordance with our results.
We obtain relatively high effective temperature $T\sim 50\,$K  for a couple of measured spectra
which means that the local exciton lifetime can be comparable to the time of energy relaxation.
The QW splitting $\Delta^{{\rm qw}}$ and g-factor of neutral acceptors were directly measured via spin-flip
Raman scattering~\cite{PhysRevB.50.2510}. Our g-factor values are comparable with those from~\cite{PhysRevB.50.2510};
the discrepancy is due to using the model fit instead of direct measurement.
We have a good agreement of $\Delta^{{\rm qw}}$ values with the data from~\cite{PhysRevB.50.2510}
if we take into account the strong fluctuation of $\Delta^{{\rm qw}}$ depending on the acceptor position
with respect to the barrier.

Such a strong dependance of $\Delta^{{\rm qw}}$ on $z$ coordinate of an individual acceptor
makes possible to establish a position of impurity in the growth direction. Lateral
coordinates of an impurity in the quantum well could be established within one-nanometer
accuracy using super-resolution optical technique which is well
developed for the single molecule spectroscopy~\cite{Thompson20022775}. An application of these
methods could provide unprecedented possibilities to establish exact atomic
coordinates of impurities inside the crystal lattice and to explore its spin and
energy structure, combining advantages of optical
spectroscopy with the ultimate spatial accuracy of the scanning tunneling microscopy.

In conclusion, we report photoluminescence measurements of excitons bound to single beryllium acceptors
in GaAs/AlGaAs QWs. In order to describe our results we use a simple theoretical model
of an acceptor-bound exciton confined in the QW. The model includes
the interparticle exchange.
The obtained parameter values of our model are in a good agreement with previously published data
and accurately describe a complex spectral signature of the single impurity in radiative recombination.

\begin{acknowledgments}
We acknowledge funding from Russian Science Foundation.
P.V.P., N.S.A., P.M.K. and A.Yu.S. were supported by project 14-42-00015 (experiments and general discussion).
I.A.K., G.V.K. and Yu.L.I. were supported by project 14-12-00255 (theory, sample growth and characterization).
\end{acknowledgments}
%

\end{document}